\documentclass[12pt]{iopart}

\usepackage{amsfonts}
\usepackage{amssymb}
\usepackage{iopams}
\usepackage{setstack}
\usepackage{graphicx}

\def\sech{\mathrm{sech}}

\newcommand{\ri}{\mathrm{i}}
\newcommand{\ud}{\mathrm{d}}
\newcommand{\ue}{\mathrm{e}}

\begin{document}

\title[Universal reductions and solitary waves of weakly nonlocal defocusing NLS equations]
{Universal reductions and solitary waves of weakly nonlocal defocusing
  nonlinear Schr\"odinger equations}

\author{G. N. Koutsokostas$^1$, T. P. Horikis$^2$, P. G. Kevrekidis$^3$,
D. J. Frantzeskakis$^1$}

\address{$^1$Department of Physics, National and Kapodistrian University of Athens,
Panepistimiopolis, Zografos, Athens 15784, Greece}

\address{$^2$Department of Mathematics, University of Ioannina, Ioannina 45110, Greece}

\address{$^3$Department of Mathematics and Statistics, University of
Massachusetts, Amherst, Massachusetts 01003-4515, USA}

\ead{horikis@uoi.gr}

\begin{abstract}

We study asymptotic reductions and solitary waves of a weakly nonlocal defocusing 
nonlinear Schr{\"o}dinger (NLS) model. The hydrodynamic form of the latter is
analyzed by means of multiscale expansion methods. To the leading-order of approximation
(where only the first moment of the response function is present), we show that 
solitary waves, in the form of dark solitons, are governed by an effective 
Boussinesq/Benney-Luke (BBL) equation, which describes
bidirectional waves in shallow water. Then, for long times, we reduce the BBL equation
to a pair of Korteweg-de Vries (KdV) equations for right- and left-going waves, and show that
the BBL solitary wave transforms into a KdV soliton. In addition, to the next order of
approximation (where both the first and second moment of the response function are present),
we find that dark solitons are governed by a higher-order perturbed KdV (pKdV) equation, which
has been used to describe ion-acoustic solitons in plasmas and water waves in the presence of 
higher-order effects. The pKdV equation is approximated by a higher-order integrable system 
and, as a result, only insubstantial changes in the soliton shape and velocity are found, 
while no radiation tails (in this effective KdV picture) are produced. 

\end{abstract}


\section{Introduction}

In many physical contexts, there appear systems featuring a spatially nonlocal nonlinearity.
For instance, in nonlinear optics, the nonlinear (intensity-dependent) correction to the
refractive index at a particular point in space, depends on the light intensity in a
certain spatial domain around this point. Such nonlocal nonlinear systems include
thermal nonlinear media, e.g., vapors \cite{su,ro} and liquid solutions \cite{tr1,tr2}
(see also the reviews \cite{krorev,mih}), plasmas \cite{pl1,pl2,pl3}, nematic liquid
crystals \cite{nem1,nem2}, dipolar Bose-Einstein condensates (BECs) \cite{beccr}, and so on.
Nonlocality has been shown to be of paramount importance on the stability and dynamics of
nonlinear waves and solitons. For instance, if the nonlocal nonlinearity is of the focusing
type, collapse can be arrested in higher-dimensions \cite{skt} (see also Ref.~\cite{krorev})
and, as a result, stable solitons exist in such settings \cite{su,ro,mih,mih6}. In the case
of defocusing nonlocal nonlinearities, dark solitons do exist \cite{attra1,dsnl1,dsnl2,dsnl3}
and may feature an attractive interaction \cite{attra1,attra2} rather than a repulsive one,
as in the case of a local nonlinearity (see the reviews \cite{yurirev,djfrev} and references
therein). Furthermore, nonlocality can suppress the transverse (``snaking'') instability
of dark solitons and the associated dispersive shock waves \cite{trillo}.

An important class of nonlocal models, relevant to the physical settings mentioned above,
is of the nonlinear Schr{\"o}dinger (NLS) type; a one-dimensional (1D) such model, expressed
in dimensionless form, is of the form:
\begin{equation}
\ri u_t+\frac{1}{2}u_{xx}
+ \sigma \left[\int_{-\infty }^{\infty }R(x'-x)|u(x',t)|^2\,\ud x' \right] u=0,
\label{nonlocal0}
\end{equation}
where $u(x,t)$ is a complex field, $R(x)$ is a positive definite function describing the
nonlocal response of the medium, and $\sigma = \pm 1$ corresponds to a focusing or a defocusing
nonlinearity (for $\sigma=+1$ and $\sigma=-1$, respectively).
Obviously, if the response function is singular, i.e., $R(x)=\delta(x)$
(where $\delta(x)$ is the Dirac $\delta$ function), then Eq.~(\ref{nonlocal0}) reduces to the usual,
completely integrable in the $(1+1)$-dimensional setting, NLS equation. Here, we are interested 
in the case where the spatial width of the response kernel $R(x)$ is small compared to 
the width of the density $|u|^2$. In such a case, nonlocality is weak and, following the analysis 
of Ref.~\cite{wnl}, it is possible to reduce Eq.~(\ref{nonlocal0}) to an effective local NLS model.
There, nonlocality is effectively described by means of
a local perturbation, in the form of a nonlinear potential term (see details below).

In this work, our scope is to study analytically the formation and dynamics of dark solitons in
the above mentioned weakly nonlocal setting, with a defocusing nonlinearity ($\sigma=-1$).
In particular, our analytical approach relies on a multiscale analysis of the hydrodynamic
form of the pertinent NLS model. This way, we derive -- at various stages of the asymptotic analysis
-- a wealth of effective nonlinear evolution equations, whose solutions are used for the construction
of approximate dark soliton solutions of the original weakly nonlocal NLS model. We are thus
able to establish interesting connections with other physical contexts.

To be more specific, at an intermediate stage of our analysis, first we derive an equation of the
form of the Boussinesq \cite{mjan} or the Benney-Luke \cite{bl} equation (hereafter, this model
will referred to as the BBL equation). The latter have been used to describe the propagation of
bidirectional waves in shallow water \cite{mjan,bl,j}, while similar Boussinesq-type models appear
in studies of waves in plasmas \cite{kar,rowl}, electrical and mechanical lattices \cite{rem},
and so on. We then use a traveling wave ansatz, and derive exact solitary wave solutions of the
BBL equation, which correspond to a weak dark soliton solution of the original NLS equation.
Next, we study the long-time behavior of the BBL equation and, similarly to the water wave problem
\cite{mjan,j}, we reduce the BBL model to a pair of Korteweg-de Vries (KdV) equations that govern
right- and left-propagating waves. We also show that if the formal perturbation parameter is
sufficiently small, then the BBL solitary wave reduces to a KdV soliton.

Finally, we use the reductive perturbation method \cite{rpm} to analyze higher-order effects arising
from the consideration of moderate widths of the response kernel. In this case, we show that
dark solitons are governed by a 5th-order perturbed KdV (pKdV) equation, which stems naturally
from the underlying Hamiltonian systems \cite{mc,men}, and is related to the first
higher-order equation in the KdV hierarchy \cite{mjacl}. This pKdV model, which
is known to describe ion-acoustic solitons \cite{kota,ich} and shallow water waves \cite{mar1}
under the influence of higher-order effects, has been studied in the context of asymptotic
integrability of weakly dispersive nonlinear wave equations \cite{fokas,kod}.
An approximate soliton solution of the derived pKdV is presented, and it is shown that it
is only slightly deformed as compared to the original KdV soliton obtained to the leading-order
of approximation. 
In addition, the perturbation theory for solitons \cite{km1,km2,yubo} is
employed in order to determine the effect of perturbation on the soliton characteristics under
the action of the higher-order effects. We find that the soliton amplitude remains unchanged,
while no radiation tails are produced during the evolution in the  higher-order KdV approximation.

The manuscript is organized as follows. In Section~2 we present the model equations,
while Section~3 is devoted to the asymptotic analysis and the presentation of the soliton
solutions. Finally, in Section~4 we present our conclusions and discuss possibilities for
future work.

\section{Model equations}

We consider a physical system, which is governed by the following dimensionless
1D nonlocal defocusing NLS equation for the unknown complex field $u(x,t)$:
\begin{equation}
\ri u_t+\frac{1}{2}u_{xx} - n(I)u=0,
\label{nonlocal}
\end{equation}
where subscripts denote partial derivatives, and the real function $n(I)$, with $I=|u(x,t)|^2$,
is given by the following convolution integral:
\begin{equation}
n(I)=\int_{-\infty }^{\infty }R(x'-x)I(x',t)\,\ud x',
\label{dn}
\end{equation}
with the kernel $R(x)$ describing the response function of the nonlocal medium. This
nonlocal model describes, in the context of optics, beam propagation in thermal media
\cite{su,ro,tr1,tr2}; in this case, $u(x,t)$ is the electric field envelope, $n$ is
the nonlinear change of the refractive index (that depends on the light intensity $I$),
while $t$ represents the propagation direction. A similar situation occurs in plasmas,
but with $n$ denoting the relative electron temperature perturbation \cite{pl1,pl2,pl3},
as well as in nematic liquid crystals, with $n$ being the perturbation of the optical
director angle from its static value due to the presence of the light field \cite{nem1,nem2}.

In all the above cases, the kernel $R(x)$ may be considered to be a real, positive definite,
localized and symmetric function \cite{comment}, obeying the normalization condition
$\int_{-\infty }^{+\infty }R(x)\ud x=1$.
A physically relevant form of the kernel,
that finds applications in all the above mentioned contexts, is:
\begin{equation}
R(x)=\frac{1}{2d}\exp\left(-\frac{|x|}{d}\right),
\label{R}
\end{equation}
where $d>0$ is a spatial scale that measures the degree of nonlocality ($d = 0$
corresponds to the limit of local nonlinearity). Then, introducing the Fourier transform
pair for a function $f(x)$ as:
\begin{eqnarray}
\hat{f}(k) = \mathcal{F}\{ f(x) \} &=& \int_{-\infty}^{\infty} f(x)\ue^{\ri kx}\,\ud x,
\nonumber \\
f(x)=\mathcal{F}^{-1}\{ \hat{f}(k) \} &=& \frac{1}{2\pi}
\int_{-\infty}^{\infty} \hat{f}(k)\ue^{-\ri kx}\,\ud k,
\nonumber
\end{eqnarray}
and taking into regard that $\hat{R}(k) = (1+d^2k^2)^{-1}$, it can readily be found that
(\ref{nonlocal}) is equivalent to the following system of coupled partial differential
equations (PDEs):
\begin{eqnarray}
\ri u_t +\frac{1}{2} u_{xx}- n u=0,
\label{example1a} \\
d^2 n_{xx}- n = -|u|^2.
\label{example1b}
\end{eqnarray}

In this work, we focus on the case where the response function is narrow as compared
to the width of the beam's intensity (the so-called weakly nonlocal limit) \cite{wnl}.
Then, $n(x,t)$ in Eq. (\ref{dn}) may be approximated appropriately, so that (\ref{nonlocal})
may be written in local form, so that such a weakly nonlocal medium may be described by a
local PDE. To do this, we express $n(x,t)$ in Fourier space and, recalling that Eq.~ (\ref{dn})
is a convolution integral, we have:
\[
\hat{n}(k,t)=\hat{R}(k)\hat{I}(k,t).
\]
Next, we expand $\hat{R}(k)$ in a Taylor series as
\[
\hat{R}(k)=\sum_{n=0}^{\infty}\frac{\hat{R}^{(n)}(0)}{n!}k^n,
\]
where $\hat{R}^{(n)}(k) \equiv d^n \hat{R}(k)/dk^n$, and using Fourier transform properties
we can find:
\begin{eqnarray}
n(x,t) &=& {\mathcal{F}^{ - 1}}\left\{ {\hat R(k)\hat I(k,t)} \right\}
= {\mathcal{F}^{-1}}\left\{ {\sum\limits_{n = 0}^\infty  {\frac{{{{\hat R}^{(n)}}(0)}}{{n!}}} {k^n}\hat I(k,t)} \right\} \nonumber\\
&=& \sum\limits_{n = 0}^\infty  {\frac{{{{\hat R}^{(n)}}(0)}}{{n!}}{\mathcal{F}^{-1}}\left\{ {{k^n}\hat I(k,t)} \right\}}
= \sum\limits_{n = 0}^\infty  {\frac{{{{\hat R}^{(n)}}(0)}}{{n!}}} {\ri^n}\frac{{{\partial ^n}}}{{\partial {x^n}}}I(x,t).
\label{dn_expand}
\end{eqnarray}
To this end, using the expansion (\ref{dn_expand}) and the properties of the Fourier transform
for $\hat{R}^{(n)}(0)$, we obtain from Eq. (\ref{nonlocal}) the following local PDE:
\begin{equation}
\ri u_t + \frac{1}{2}u_{xx}-\left[ \sum_{n = 0}^\infty
a_n \frac{{{\partial ^n}}}{{\partial {x^n}}}\left(|u{|^2}\right) \right] u = 0,
\label{nonlocal2}
\end{equation}
where
\begin{equation}
{a_n} = 
{\frac{{{{( - 1)}^n}}}{{n!}}} \left[ {\int\limits_{-\infty}^\infty {{x^n}R(x)dx} } \right].
\label{an}
\end{equation}
Note that if $n=0$ (corresponding to the singular $\delta$-function kernel) then
$\hat{R}^{(0)}=\hat{R}(0)=\int_{-\infty}^{\infty} R(x)\,\ud x=1$, i.e., $a_0=1$,
and Eq. (\ref{nonlocal2}) reduces to the cubic NLS equation, while for
$n \ne 0$,
given the symmetric nature of $R$, we have:
$$a_{2n}>0, \quad {\rm and} \quad a_{2n+1}=0, \quad \forall n \in \mathbb{N}.$$
For instance, in the case of the kernel
$R(x)$ given by Eq.~(\ref{R}), it is straightforward to find that:
\[
\int_{-\infty }^{+\infty}  {x^n}R(x)dx = \frac{1}{2d} \int_{-\infty }^{+\infty}{{x^n}{\ue^{ - |x|/d}}\,\ud x
= } \left\{ {\begin{array}{l}
  {{d^n}n!,\;n\,\mathrm{even}} \\
  {0,\;n\, \mathrm{odd}}
\end{array}} \right.
\]
and hence $a_{2n}=d^{2n}$.

According to the above discussion, in the case of a response kernel of sufficiently small width
compared to the width of the intensity $I(x,t)\equiv|u(x,t)|^{2}$, we may use, to a first
approximation, $n(I)\approx I+a_2 \partial _{x}^{2}I$, and find that Eq.~(\ref{nonlocal2}) is
reduced to the following modified NLS equation:
\begin{equation}
\ri u_{t}+\frac{1}{2}u_{xx} - \left( |u|^{2}+a_2 \partial_{x}^{2}|u|^{2} \right)u=0,
\label{nlnls}
\end{equation}
with the parameter $a_2$ characterizing the nonlocality. The above equation has been studied in
nonlinear optics \cite{wnl}, as well as (in the case of a focusing nonlinearity) in plasma physics,
where the parameter $a_2$ may take both positive and negative values \cite{dav}, and the continuum
limit of discrete molecular structures \cite{wang}. It has also been shown that Eq.~(\ref{nlnls})
possesses stable soliton solutions, bright or dark for a focusing or a
defocusing nonlinearity (i.e., a $+$ or a $-$ sign in front of the
parenthesis
in Eq.~(\ref{nlnls}))
respectively \cite{wnl}.

Here, we are interested in studying the role of higher-order effects on dark soliton dynamics,
which may be accounted for by the inclusion of higher-order terms in the Taylor expansion of
$n(x,t)$. Thus, below, we will use multiscale expansion methods to study dark solitons of
Eq.~(\ref{nonlocal2}) up to the following level of approximation:
$n(I)\approx I+a_2 \partial _{x}^{2}I+a_4 \partial_{x}^{4}I $.
In this case, Eq.~(\ref{nonlocal2}) is obviously reduced to the following higher-order
modified NLS equation:
\begin{equation}
\ri u_{t}+\frac{1}{2}u_{xx}-\left( |u|^{2}+a_2 \partial_{x}^{2}|u|^{2}
+a_4 \partial_{x}^{4}|u|^{2} \right) u=0.
\label{nlnls2}
\end{equation}
In the next Section, we will use multiscale expansion methods and derive universal models the 
soliton solutions of which will then be used to obtain approximate dark soliton solutions
of Eq.~(\ref{nlnls2}).

\section{Asymptotic analysis}

\subsection{The cw solution and its stability}

To start our analysis, first we introduce the Madelung transformation
$$u(x,t)=u_{0}\rho^{1/2}(x,t) \exp[\ri\phi(x,t)],$$
(where $u_0$ is an arbitrary real constant), and derive from Eq.~(\ref{nonlocal2}) the following
system of two coupled PDEs for the amplitude $\rho$ and phase $\phi$,
\begin{eqnarray}
\phi _{t}+u_{0}^{2}\sum_{n = 0}^\infty a_{2n} \frac{{{\partial^{2n}}}}{{\partial {x^{2n}}}}\rho
+\frac{1}{2}\phi _{x}^{2}
-\frac{1}{2}\rho ^{-1/2}\left( \rho ^{1/2}\right) _{xx}=0,
\label{real}
\\
\rho_{t}+\left( \rho \phi _{x}\right) _{x}=0.
\label{imag}
\end{eqnarray}
Obviously, the system~(\ref{real})-(\ref{imag}) possesses a simple
homogeneous solution, namely:
\begin{equation}
\rho=1, \quad \phi=-u_{0}^{2}t,
\label{cw}
\end{equation}
which corresponds to the continuous-wave (cw) solution $u=u_{0} \exp(-\ri u_{0}^{2}t)$
of Eq.~(\ref{nonlocal2}). Since below we will seek dark soliton
solutions of Eq.~(\ref{nonlocal2}) on top of this cw background, it is necessary to investigate
the stability of the cw solution. To do so, we assume that $\rho=1+\Delta\rho$,
$\phi=-u_{0}^{2}z+\Delta\phi$, where the perturbations $\Delta\rho$, $\Delta\phi$
(with $|\Delta\rho|\ll 1$, $|\Delta\phi|\ll 1$) behave like $\exp[\ri(kx-\omega t)]$; this way,
Eqs.~(\ref{real})-(\ref{imag}) lead to the following dispersion relation for the perturbations'
frequency $\omega$ and wavenumber $k$:
\begin{equation}
\omega^{2}= k^{2}\left[ c^2 \sum_{n = 0}^\infty a_{2n} (ik)^{2n}+\frac{1}{4}k^2\right],
\label{dp0}
\end{equation}
where
\begin{equation}
c^2 = u_{0}^{2},
\label{ss}
\end{equation}
is the speed of the small-amplitude (linear) waves propagating on top of the cw background,
the so-called ``speed of sound''. Equation~(\ref{dp0}) shows that the cw is modulationally stable,
i.e., $\omega \in \mathbb{R}~\forall k\in \mathbb{R}$, provided that
$\sum_{n = 0}^\infty a_{2n} (ik)^{2n}>0$. Obviously this occurs for $n=0$ (corresponding
to the local NLS with the defocusing nonlinearity). On the other hand, for $n=1$ the cw is
modulationally stable as long as the parameter $\alpha$, defined as:
\begin{equation}
\alpha = 1-4u_{0}^{2} a_2,
\label{p}
\end{equation}
is positive. This condition is fulfilled if, for a fixed nonlocality parameter $a_2$,
the cw background intensity $u_{0}^{2}$ does not exceed a critical value $I_{0}^{(cr)}$,
i.e., $u_{0}^{2} \leq I_{0}^{(cr)}\equiv (4a_2 )^{-1}$, in accordance to the analysis of
Ref.~\cite{minl}; the same holds if, for a fixed cw intensity $u_{0}^{2}$, the nonlocality
parameter $a_2$ does not exceed the critical value $a_2^{(cr)}$, i.e.,
$a_2 \leq a_2^{(cr)} \equiv (4u_{0}^{2})^{-1}$.
In addition, for $n=2$, the dispersion relation~(\ref{dp0}) becomes:
\begin{equation}
\omega^{2}=c^2 k^{2}\left( 1+\frac{\alpha}{4 c^{2}}k^{2}+a_4 k^{4}\right).
\label{dp}
\end{equation}
Notice that in the case of the kernel~(\ref{R}),
the series in Eq.~(\ref{dp0}) converges to $(1+d^2 k^2)^{-1}$ and, as a result, Eq.~(\ref{dp0})
takes the form:
\begin{equation}
\omega^2=\frac{c^2 k^2}{1+d^2 k^2}+\frac{1}{4}k^4,
\label{dr}
\end{equation}
which is the dispersion relation of the nonlocal model~(\ref{example1a})-(\ref{example1b}).
Obviously, in this fully nonlocal case, the cw background is always stable.

Before proceeding with the asymptotic analysis of the weakly nonlocal NLS model, it is worth
mentioning the following. For right-going waves, the dispersion relation~(\ref{dp}) becomes
$\omega=ck \left[1+(\alpha/4c^2)k^2+a_4 k^4\right]^{1/2}$, and in the long-wavelength limit
($k \ll 1$), it can be reduced to the form:
$$\omega \approx ck + \frac{\alpha}{8c} k^3+\left(\frac{1}{2}ca_4
-\frac{\alpha^2}{128c^3} \right)k^5.$$
Then, using $\omega \mapsto \ri\partial_t$ and $k \mapsto -\ri\partial_x$ to
revert to the corresponding (linear) PDE, and introducing a reference frame moving 
with velocity $c$, i.e., $x \mapsto x-ct$, it can be found that the linear PDE for 
a field $Q(x,t)$ associated to the above dispersion relation is:
\begin{equation}
Q_t-\frac{\alpha}{8c} Q_{xxx}
+ \left(\frac{1}{2}ca_4 -\frac{\alpha^2}{128c^3} \right)Q_{xxxxx}=0.
\label{lpde}
\end{equation}
The above PDE has the form of a linearized 5th-order KdV equation. The full nonlinear
version will be derived below in Section~\ref{3D}.

\subsection{The Boussinesq / Benney-Luke equation and the solitary wave solution}

We first consider an intermediate stage of the asymptotic analysis, and seek 
solutions of Eqs.~(\ref{real})-(\ref{imag}) in the form of the following asymptotic expansions:
\begin{eqnarray}
\phi=-u_0^2 t + \epsilon^{1/2}\Phi(X,T),
\quad
\rho= 1+\sum_{j=1}^{\infty}\epsilon^j \rho_j(X,T),
\label{asex}
\end{eqnarray}
where $0<\epsilon \ll1$ is a formal small parameter that sets the soliton's amplitude
[i.e., below, we will find soliton solutions valid up to order $\mathcal{O}(\epsilon)$].
Here, it is assumed that the phase $\Phi$ and
amplitudes $\rho_j$ are unknown real functions of the slow variables:
\begin{equation}
X=\epsilon^{1/2}x, \quad T=\epsilon^{1/2}t.
\label{slow}
\end{equation}
Substituting the expansions~(\ref{asex}) into Eqs.~(\ref{real})-(\ref{imag}), and equating
terms of the same order in $\epsilon$, we obtain the following results. First, Eq.~(\ref{real})
reads:
\begin{eqnarray}
\Phi_T + u_0^2 \rho_1 +\epsilon \left(u_0^2 \rho_2
+\frac{1}{2}\Phi_X^2-\frac{1}{4}\alpha \rho_{1XX} \right)
= \mathcal{O}(\epsilon^2),
\label{Phi0}
\end{eqnarray}
while Eq.~(\ref{imag}) yields, at $\mathcal{O}(\epsilon^{3/2})$ and
$\mathcal{O}(\epsilon^{5/2})$, the following equations respectively:
\begin{eqnarray}
\rho_{1T}+\Phi_{XX}=0,
\label{i1} \\
\rho_{2T}+\left(\rho_1 \Phi_{X}\right)_X=0.
\label{i2}
\end{eqnarray}
Differentiating Eq.~(\ref{Phi0}) once with respect to $T$, and using $\rho_{1T}=-(1/u_0^2)\Phi_{TT}$
[from the leading-order part of Eq.~(\ref{Phi0})] as well as Eqs.~(\ref{i1})-(\ref{i2}), we
eliminate the functions $\rho_1$ and $\rho_2$ from the resulting equation, and arrive at
the following equation for $\Phi$:
\begin{eqnarray}
\Phi_{TT}-c^2\Phi_{XX}+\epsilon \left[ \frac{1}{4}\alpha \Phi_{XXXX}
+\frac{1}{2}\left(\Phi_X^2\right)_T+\left(\Phi_T \Phi_X\right)_X \right]
=\mathcal{O}(\epsilon^2).
\label{bbl}
\end{eqnarray}
At the leading-order, Eq.~(\ref{bbl}) is a 2nd-order linear wave equation, with the wave velocity
$c$ given by Eq.~(\ref{ss}), as found above. On the other hand, at order $\mathcal{O}(\epsilon)$,
Eq.~(\ref{bbl}) features a fourth-order dispersion and quadratic nonlinear terms, thus resembling
the $(1+1)-$dimensional variants of the Boussinesq \cite{mjan} or Benney-Luke \cite{bl} equations.

It is now possible to derive an exact solitary wave solution to the above 
Boussinesq/Beney-Luke (BBL) model, Eq.~(\ref{bbl}). This can be done upon seeking traveling 
wave solutions of the form:
\begin{equation}
\Phi=\Phi(s), \quad s=X-vT,
\end{equation}
where $v$ is the velocity of the traveling wave. Assuming vanishing boundary conditions
for $\Phi'\equiv d\Phi/ds$, i.e., $\Phi' \rightarrow 0$ as $s\rightarrow \pm \infty$, 
we substitute in Eq.~(\ref{bbl}) and obtain the following 3d-order ordinary differential 
equation (ODE):
\begin{equation}
\frac{1}{4}\epsilon \alpha \Phi''' + (v^2-c^2)\Phi' -\frac{3}{2}\epsilon v \Phi'^2=0,
\label{ode0}
\end{equation}
where primes denote derivatives with respect to $s$. Next, we assume that the unknown field
$\rho_1$ also depends on $s$, i.e., $\rho_1=\rho_1(s)$, with $\rho_1(s) \rightarrow 0$ as
$s\rightarrow \pm \infty$. Then, we may use $\Phi_{T}=-u_0^2 \rho_{1}$
from the leading-order part of Eq.~(\ref{Phi0}) and obtain the auxiliary equation
$\Phi'=(c^2/v)\rho_1$. Substituting the latter equation into Eq.~(\ref{ode0}), we derive
the following 2nd-order ODE for $\rho_1$:
\begin{equation}
\rho_1''+\frac{4}{\epsilon \alpha}(v^2-c^2)\rho_1 - \frac{6c^2}{\alpha}\rho_1^2=0,
\label{ode1}
\end{equation}
Equation~(\ref{ode1}) can be seen as the equation of motion of a unit mass particle in the
presence of the effective potential
$V(\rho_1)=(2/\epsilon \alpha)(v^2-c^2)\rho_1^2-(2c^2/\alpha)\rho_1^3$.
We assume that $v^2-c^2<0$, i.e., we focus on traveling waves moving with a velocity
smaller than the speed of sound (subsonic waves). Then, a simple analysis shows that,
in this case, there exists a hyperbolic fixed point,
at $\rho_1= 0$ [corresponding to the global maximum of $V(\rho_1)$], and an elliptic fixed
point, at $\rho_1=(2/3\epsilon c^2)(v^2-c^2)$ [corresponding to the global minimum of $V(\rho_1)$].
In the phase plane of the system, associated to the hyperbolic fixed point that corresponds
to zero energy $E$, i.e., for $(1/2)\rho_1'^2+V(\rho_1)=E=0$, there exists a homoclinic orbit
(separatrix). The latter, is a trajectory of infinite period, which corresponds to a solution
decaying at infinity, i.e., a solitary wave with vanishing asymptotics (Note that if $v^2-c^2>0$
the hyperbolic fixed point would become an elliptic one, and vice versa, and as a result the
corresponding solitary wave would not asymptote to zero, as per our assumption above).
It is then straightforward to find that this solitary wave solution
can be expressed in the following explicit form:
\begin{equation}
\rho_1=-\frac{1}{\epsilon}\left(1-\frac{v^2}{c^2}\right)
{\rm sech}^2\left[\frac{c}{\sqrt{\epsilon \alpha}}
\left(1-\frac{v^2}{c^2}\right)^{1/2} (X-vT-X_0)\right],
\label{esw}
\end{equation}
where $X_0$ is an arbitrary constant that sets the initial soliton location.
It can now readily be seen that the amplitude of the soliton is $\mathcal{O}(\epsilon^{-1})$,
while, generally, the condition $|{\rm max}(\rho_j)|=\mathcal{O}(1)
~\forall j\in \mathbb{N}$ should hold, as implied by the asymptotic expansion of $\rho$ in
Eq.~(\ref{asex}). Hence, in order for the solution to be meaningful, 
i.e., the soliton amplitude is $\mathcal{O}(1)$, we assume that 
the (arbitrary so far) velocity $v$ is sufficiently close to $c$, namely the following
condition holds:
\begin{equation}
v^2/c^2 = 1-\epsilon \mu^2,
\label{mu}
\end{equation}
where $\mu$ is a $\mathcal{O}(1)$ parameter (recall that $v^2-c^2<0$). Notice that, under 
this assumption, we detune from the sonic limit and explicitly consider solely solitary 
waves (which are gray solitons in the original model -- see below) within this region. 
Then, the solution~(\ref{esw}) becomes:
\begin{equation}
\rho_1= -\mu^2 {\rm sech}^2\left[\frac{c\mu}{\sqrt{\alpha}}(X-c\sqrt{1-\epsilon \mu^2}T-X_0)\right],
\label{sol0}
\end{equation}
and thus, an approximate solution [valid up to $\mathcal{O}(\epsilon)$] of Eq.~(\ref{nonlocal2})
is of the form:
\begin{eqnarray}
\hspace{-2cm}u(x,t) \approx u_0 \left[1-\epsilon \mu^2 {\rm sech}^2 (\epsilon^{1/2}\theta) \right]^{1/2}
\exp\left[-\ri u_{0}^{2}t + \frac{\ri \epsilon^{1/2}c}{\sqrt{1-\epsilon \mu^2}}
\tanh(\epsilon^{1/2}\theta) \right],
\label{apsol1} \\
\hspace{-2cm}\theta=\frac{c\mu}{\sqrt{\alpha}}\left(x-c \sqrt{1-\epsilon \mu^2} t-x_0\right),
\label{apsola}
\end{eqnarray}
where we have used the equation $\Phi'=(c^2/v)\rho_1$ to derive the phase of the solution.
It is clear that Eq.~(\ref{apsol1}) represents a density dip on top of the cw background,
with a phase jump across the density minimum, and hence it is a dark soliton.

Notice that, in our analysis above, the velocity $v$ may be either positive or 
negative and, therefore, propagation of either right- or left-going waves is allowed, 
as should be expected from the bidirectional BBL equation. Below we will show that, indeed, 
and similarly to the water wave problem \cite{mjan}, the far-field of Eq.~(\ref{bbl})  
is a pair of two KdV equations, for right- and left-going waves, and that the solitary 
wave~(\ref{sol0}) is accordingly reduced to right-going KdV soliton.

\subsection{The KdV equation and the soliton solution}

We now proceed to obtain the far-field equations stemming from the
BBL model~(\ref{bbl}) by means of a multiscale asymptotic
expansion method.  
We seek solutions of Eq.~(\ref{bbl}) in the form of the asymptotic expansion:
\begin{equation}
\Phi = \sum_{j=0}^{\infty} \epsilon^j \Phi_j,
\label{expphi}
\end{equation}
where the unknown functions $\Phi_j$~($j=0,1,2,\ldots$) depend on the new variables:
\begin{equation}
\xi = X-cT, \quad \eta=X+cT, \quad \tau=\epsilon T.
\label{nsv}
\end{equation}
Substituting Eq.~(\ref{expphi}) into Eq.~(\ref{bbl}), we obtain the following results.
First, at the leading-order, $\mathcal{O}(1)$:
\begin{equation}
4c^2\Phi_{0\xi\eta}=0,
\label{losp}
\end{equation}
which implies that $\Phi_0$ is a superposition of a right-going wave, $\Phi_0^{(R)}$,
depending on $\xi$, and a left-going one, $\Phi_0^{(L)}$, depending on $\eta$, namely:
\begin{equation}
\Phi_{0}=\Phi_0^{(R)}(\xi)+\Phi_0^{(L)}(\eta).
\label{rlcarsp}
\end{equation}
Second, at order $\mathcal{O}(\epsilon)$:
\begin{eqnarray}
4c^2\Phi_{1\xi\eta} = &-&c\left(\Phi_{0\xi\xi}^{(R)}\Phi_{0\eta}^{(L)}
-\Phi_{0\xi}^{(R)}\Phi_{0\eta\eta}^{(L)} \right) \nonumber \\
&+&\left(
-2c\Phi_{0\tau}^{(R)} +\frac{\alpha}{4}\Phi_{0\xi\xi\xi}^{(R)}
-\frac{3c}{2}\Phi_{0\xi}^{(R)2}\right)_{\xi}
\nonumber \\
&+&\left(2c\Phi_{0\tau}^{(L)}
+\frac{\alpha}{4}\Phi_{0\eta\eta\eta}^{(L)}
+\frac{3c}{2}\Phi_{0\eta}^{(L)2}
\right)_{\eta}.
\label{phicarsp}
\end{eqnarray}
It is clear that upon integrating Eq.~(\ref{phicarsp}) in $\xi$ or $\eta$,
the terms in parentheses in the right-hand side of this equation are secular,
because are functions of $\xi$ or $\eta$ alone, and not both. Removal of these terms
leads to two uncoupled nonlinear evolution equations for $\Phi_0^{(R)}$ and $\Phi_0^{(L)}$.
Furthermore, employing $\Phi_{TT}=-u_0^2 \rho_{1T}=-c^2 \rho_{1T}$
%
%
from the leading-order
part of Eq.~(\ref{Phi0}) it can readily be seen that the amplitude
$\rho_1$ can also be decomposed to a left- and a right-going wave,
i.e., $\rho_1 = \rho_1^{(R)}+\rho_1^{(L)}$, with
\begin{equation}
\Phi_{0\xi}^{(R)}=c\rho_1^{(R)}, \quad \Phi_{0\eta}^{(L)}=-c\rho_1^{(L)}.
\end{equation}
Then, using the above expressions, the equations for $\Phi_0^{(R)}$ and $\Phi_0^{(L)}$
stemming from Eq.~(\ref{phicarsp}) yield the following two uncoupled KdV equations for
$\rho_1^{(R)}$ and $\rho_1^{(L)}$:
\begin{eqnarray}
\rho_{1\tau}^{(R)} -\frac{\alpha}{8c}\rho_{1\xi\xi\xi}^{(R)}
+\frac{3c}{2}\rho_{1}^{(R)} \rho_{1\xi}^{(R)}=0,
\label{kdvr} \\
\rho_{1\tau}^{(L)}
+\frac{\alpha}{8c}\rho_{1\eta\eta\eta}^{(L)}
-\frac{3c}{2}\rho_{1}^{(L)} \rho_{1\eta}^{(L)}=0.
\label{kdvl}
\end{eqnarray}
We have thus shown that, indeed, the far field of Eq.~(\ref{bbl}) is a pair of KdV equations
for a left- and a right-going wave. We will also show that the KdV soliton is directly
connected with the solitary wave solution~(\ref{sol0}) of the BBL equation. To do this,
let us consider the right-going wave, and write down the soliton solution of
the KdV Eq.~(\ref{kdvr}):
\begin{equation}
\rho_{1}(\xi,\tau)=-\frac{\kappa^{2}\alpha}{c^{2}}{\rm sech}^{2}Z,
\quad
Z= \kappa\left(\xi+\frac{\kappa^{2}\alpha}{2c}\tau - \xi_0\right),
\label{kdvsoli}
\end{equation}
where $\kappa$ is an arbitrary $\mathcal{O}(1)$ parameter. When expressed in terms of the original
variables, the above KdV soliton reads:
\begin{equation}
\rho_{1}(x,t)=-\frac{\kappa^{2}\alpha}{c^{2}}{\rm sech}^{2}
\left\{ \epsilon^{1/2}\kappa \left[x-c\left(1-\epsilon \frac{\kappa^{2}\alpha}{2c^2}\right)t 
- x_0\right] \right\},
\label{kdvsoli2}
\end{equation}
showing that the velocity of the KdV soliton is $v_s=c[1-\epsilon(\kappa^{2}\alpha)/(2c^2)]$.
On the other hand, returning back to the solitary wave~(\ref{sol0}), it is observed that if the
free parameter $\mu$ is taken to be such that:
\begin{equation}
\mu^2=\kappa^2\alpha/c^2,
\label{cmu}
\end{equation}
the solitary wave~(\ref{sol0}) can be expressed in terms of the original variables as:
\begin{equation}
\rho_1= -\frac{\kappa^{2}\alpha}{c^{2}} {\rm sech}^2
\left[\epsilon^{1/2}\kappa \left(x-c\sqrt{1-\epsilon\frac{\kappa^{2}\alpha}{c^2}}~t 
- x_0\right)\right],
\label{sol02}
\end{equation}
and, hence, the solitary wave velocity is $V_s=c\sqrt{1-\epsilon(\kappa^{2}\alpha)/c^2}$.
It is now clear that for sufficiently small $\epsilon$ we may use the approximation
$V_s \approx v_s=c[1-\epsilon(\kappa^{2}\alpha)/(2c^2)]$, showing that
the solitary wave~(\ref{sol0}) transforms into the KdV soliton~(\ref{kdvsoli2}).
The latter, gives rise to an approximate dark soliton solution of Eq.~(\ref{nonlocal2}),
similar to that in Eq.~(\ref{apsol1}), but with $\mu^2$ given by Eq.~(\ref{cmu})
and with the substitution $\sqrt{1-\epsilon \mu^2} \mapsto 1-\epsilon \mu^2/2$.

At this point, we should mention that, in the case of the weakly nonlocal system we consider here,
the presented approximate solutions are always dark solitons, due to the condition $\alpha>0$
following from the requirement of the stability of the cw background state. Nevertheless, in the
fully nonlocal system characterized by the kernel~(\ref{R}) [see
Eqs.~(\ref{example1a})-(\ref{example1b})], the cw is always modulationally stable as mentioned
above. As a result, the parameter $\alpha$ may also take negative values and, thus, the nonlocal
system, also possesses approximate {\it antidark} soliton solutions (i.e., density humps rather
than dips on top of the cw background), as predicted in Refs.~\cite{tph,prl,prsa}. Obviously,
these are supersonic structures (here, $V_s>c$) which can only be found in the strongly nonlocal 
regime, and can not be supported in the weakly nonlocal case under consideration.


\subsection{Reductive perturbation method and higher-order effects}
\label{3D}

The formal derivation of the KdV equation~(\ref{kdvr}) from the BBL model~(\ref{bbl})
for long times, and particularly at the scales of Eq.~(\ref{nsv}), suggest that the KdV
equation (e.g., for the right-going wave) can also be obtained directly from the hydrodynamic
equations~(\ref{real})-(\ref{imag}). This can be done upon employing the reductive perturbation
method (RPM) \cite{rpm}. In the framework of the RPM, as we will see, it is straightforward
to take into regard higher-order effects [namely, the term $a_4 \rho_{xxxx}$ in Eq.~(\ref{real})]
and thus derive an effective higher-order KdV equation describing dark solitons in
weakly nonlocal media. 

We start by seeking solutions of Eqs.~(\ref{real})-(\ref{imag}) in the form of the asymptotic expansions
\begin{eqnarray}
\phi=-u_0^2 t + \sum_{j=0}^{\infty}\epsilon^{j+1/2} \phi_j(\xi,\tau),
\quad
\rho= 1+\sum_{j=1}^{\infty}\epsilon^j \rho_j(\xi,\tau),
\label{asex2}
\end{eqnarray}
where the unknown functions $\rho_{j}$ and $\phi_{j}$ ($j \in \mathbb{N}$) depend
on the stretched coordinates:
\begin{equation}
\xi=\epsilon^{1/2}\left(x-ct\right), \quad \tau=\epsilon^{3/2}t,
\label{strcoor}
\end{equation}
where $c$ is speed of sound [see Eq.~(\ref{ss})]. Substituting the
expansions (\ref{asex2}) into Eqs.~(\ref{real})-(\ref{imag}), and using the
variables (\ref{strcoor}), we obtain a hierarchy of coupled equations, which are to
be solved order by order in $\epsilon$ [note that, as before, both parameters $a_2$ and $a_4$
are assumed to be of order $\mathcal{O}(1)$]. Particularly,
to the leading order, i.e., to orders $\mathcal{O}(\epsilon)$ and $\mathcal{O}(\epsilon^{3/2})$,
Eqs.~(\ref{real})-(\ref{imag}) respectively lead to the following linear equations,
\begin{eqnarray}
c\phi_{1\xi}-u_{0}^{2}\rho_{1}&=&0,
\label{lin0}\\
c\rho_{1\xi}-\phi_{1\xi\xi}&=&0.
\label{lin}
\end{eqnarray}
The compatibility condition of Eqs.~(\ref{lin}) is the algebraic
Eq.~(\ref{ss}). To the next order, namely to order
$\mathcal{O}(\epsilon^{2})$ and $\mathcal{O}(\epsilon^{5/2})$, Eqs.~(\ref{real})-(\ref{imag})
respectively read
\begin{eqnarray}
c\phi_{2\xi}-u_{0}^{2}\rho_{2}&=&
\phi_{1\tau}-\frac{\alpha}{4}\rho_{1\xi\xi}+\frac{1}{2}\phi_{1\xi}^{2},
\label{sec1}
\\
c\rho _{2\xi}-\phi_{2\xi\xi}&=&\rho_{1\tau}+\left( \rho_{1}\phi_{1\xi}\right)_{\xi},
\label{sec2}
\end{eqnarray}
with $\alpha$ given in Eq.~(\ref{p}). Using the second of Eqs.~(\ref{lin}), the unknown function
$\phi_{1}$ is expressed by means of $\rho_{1}$, i.e.,
\begin{equation}
\phi_{1\xi}=c\rho_1.
\label{f1}
\end{equation}
Then, the compatibility conditions of Eqs.~(\ref{sec1})-(\ref{sec2}) are determined,
once Eq.~(\ref{sec1}) is differentiated with respect to $\xi$, Eq.~(\ref{sec2}) is multiplied
by $c$, and the resulting equations are added. This way, it is found that the
compatibility condition at this order is the KdV equation~(\ref{kdvr})
for the unknown amplitude function $\rho_{1}$, in accordance with the analysis of
the previous Section.

Notice that in the present order of approximation, there is no contribution from the term
$a_4 u \partial_{x}^{4}|u|^{2}$ in Eq.~(\ref{nonlocal2}), since the corresponding 4th-order
derivative term $a_4 \rho_{xxxx}$ in Eq.~(\ref{real}) is of higher-order. However, this term
contributes in the next order of approximation. Indeed, proceeding to the next order, 
namely to $\mathcal{O}(\epsilon^{3})$ and $\mathcal{O}(\epsilon^{7/2})$, 
Eqs.~(\ref{real})-(\ref{imag}) respectively lead to the following  equations,
\begin{eqnarray}
c\phi_{3\xi}-u_{0}^{2}\rho_{3} &=&
\phi_{2\tau}-\frac{\alpha}{4}\rho_{2\xi\xi}
+c^{2}a_4 \rho_{1\xi\xi\xi\xi}
+\phi_{1\xi}\phi_{2\xi}
+\frac{1}{8}\rho_{1\xi}^2+\frac{1}{4}\rho_{1}\rho_{1\xi\xi},
\label{thir1}
\\
c\rho_{3\xi}-\phi_{3\xi\xi}&=&
\rho_{2\tau}+\left(\rho _{1}\phi _{2\xi}\right)_{\xi}
+\left( \rho_{2}\phi _{1\xi}\right)_{\xi}.
\label{thir2}
\end{eqnarray}
The compatibility conditions of Eqs.~(\ref{thir1})-(\ref{thir2}), can also be obtained upon
following the procedure described above. In particular, first, Eq.~(\ref{thir1}) is differentiated
with respect to $\xi$, Eq.~(\ref{thir2}) is multiplied by $c$, and the resulting equations are
added. Second, we use Eq.~(\ref{f1}) to express $\phi_{1\xi}$ in terms of $\rho_1$, as well as
Eq.~(\ref{sec2}) to express $\phi_{2\xi}$ in terms of $\rho_2$ and $\rho_1$, i.e.,
\begin{equation}
\phi_{2\xi}=c\rho_2-c\rho_1^2-\int \rho_{1\tau}{\rm d}\xi,
\end{equation}
where integration constants are equal to zero due to the boundary conditions. This way, we obtain
from Eqs.~(\ref{thir1})-(\ref{thir2}) the following equation, which involves solely the fields
$\rho_1$ and $\rho_2$:
\begin{eqnarray}
\hspace{-1cm}{\rho _{2\tau }} &-& \frac{\alpha }{{8c}}{\rho _{2\xi \xi \xi }}
+ \frac{{3c}}{2}{({\rho _1}{\rho _2})_\xi } + \frac{1}{{2c}}\int {{\rho _{1\tau \tau }}d\xi }
+ {\rho _{1\xi }}\int {{\rho _{1\tau }}d\xi }  + \frac{3}{4}{(\rho _1^2)_\tau }
+ \frac{c}{2}{(\rho _1^3)_\xi }
\nonumber\\
\hspace{-1cm}&+& \frac{{{a_2}c}}{2}{(\rho _{1\xi }^2)_\xi } 
- \frac{\alpha }{{8{c^2}}}{\rho _{1\xi \xi \tau }}
- \frac{{ 1 - 8{a_2}{c^2}}}{{8c}}{\rho _1}{\rho _{\xi \xi \xi }}
+ \frac{{{a_4}c}}{2}{\rho _{1\xi \xi \xi \xi \xi }} = 0
\label{t0}
\end{eqnarray}
To this end, we multiply Eq.~(\ref{t0}) by $\epsilon$, and add it to
the KdV of Eq.~(\ref{kdvr}). Then, introducing the combined amplitude function
\begin{equation}
q=\rho_{1}+\epsilon \rho_{2},
\end{equation}
we obtain the following nonlinear evolution equation for the field $q(\xi,\tau)$:
\begin{eqnarray}
q_{\tau}-\frac{\alpha}{8c}q_{\xi\xi\xi}+\frac{3c}{2}qq_{\xi}+\epsilon P(q)= O(\epsilon^{2}),
\label{pKdV} \\
P(q) \equiv c_1 q^2 q_\xi+ c_2 q_\xi q_{\xi\xi} + c_3 q q_{\xi\xi\xi} +c_4 q_{\xi\xi\xi\xi\xi}.
\label{pofq}
\end{eqnarray}
Notice that the terms involving $\rho_{1\tau}$ in Eq.~(\ref{t0}) have been evaluated by substituting
$\rho_{1\tau}$ from the KdV, Eq.~(\ref{kdvr}). The coefficients $c_j$ ($j=1,2,3,4$) in
Eq.~(\ref{pofq}) are given by:
\begin{eqnarray}
\hspace{-2cm}
c_1=-\frac{3}{8}c, \quad
c_2=\frac{1}{4c}\left(1+\frac{5}{8}\alpha \right), \quad
c_3=\frac{1}{8c}\left(1-\frac{1}{2}\alpha \right), \quad
c_4=\frac{1}{2}ca_4 -\frac{\alpha^2}{128c^3}.
\label{cj}
\end{eqnarray}
It is readily seen that Eq.~(\ref{pKdV}) has the form of a 5th-order perturbed KdV (pKdV) equation.
It is worth observing that Eq.~(\ref{pKdV}) is reduced to the unperturbed KdV Eq.~(\ref{kdvr})
in the limit of $\epsilon=0$, while its linearized version is identical to Eq.~(\ref{lpde}).

The 5th-order pKdV equation~(\ref{pKdV}) has attracted attention, as a model 
describing the evolution of steeper waves, with shorter wavelengths than in the KdV model. 
As such, this equation has been used to describe solitons in plasmas \cite{kota,ich} 
and shallow water waves \cite{mar1} in the presence of higher-order effects, and
as a generic model that can be used to explain soliton emergence in experiments from arbitrary
initial data, even when the Hamiltonian perturbations are quite large \cite{mc,men}. 

Additionally, an extended KdV equation, similar in form to Eq.~(\ref{pKdV}), is related to 
the first higher-order equation in the KdV hierarchy \cite{mjacl}. 
In particular, using the transformations $\tau \mapsto -8c/\alpha \tau$ and 
$q \mapsto -(\alpha/2c^2)q$, Eq.~(\ref{pKdV}) reduces [up to $O(\epsilon)$] to the form: 
$$q_{\tau}+q_{\xi\xi\xi}+6qq_{\xi}
+\epsilon \left(\tilde{c}_1 q^2 q_\xi+ \tilde{c}_2 q_\xi q_{\xi\xi} 
+ \tilde{c}_3 q q_{\xi\xi\xi} +\tilde{c}_4 q_{\xi\xi\xi\xi\xi} \right)=0,$$ 
where 
$$\tilde{c}_1= -(2\alpha/c^3)c_1, \quad \tilde{c}_{2,3}= (4/c)c_{2,3}, \quad 
\tilde{c}_4= -(8c/\alpha)c_4,$$
where $c_j$ are given by Eq.~(\ref{cj}). In this case, 
the first higher-order equation in the KdV hierarchy is characterized by 
the following values of the coefficients:
$$\tilde{c}_1=1, \quad \tilde{c}_2=\frac{2}{3}, \quad 
\tilde{c}_3=\frac{1}{3}, \quad \tilde{c}_4=\frac{1}{30}.$$
Obviously, in our case, Eq.~(\ref{pKdV}) never falls in that integrable limit.
In the more general nonintegrable case, the pKdV equation with 
arbitrary $c_j$ has been studied in the context of asymptotic integrability of 
weakly dispersive and nonlinear wave equations \cite{fokas,kod}. In this context, it was 
shown that there exist asymptotic transformations \cite{fokas,kod,kra,mar3} that 
reduce the pKdV to the KdV equation. Thus, an approximate [valid up
to $O(\epsilon)$] soliton solution of Eq.~(\ref{pKdV}) can be found, which has the form of the
traditional KdV soliton with a velocity-shift and a bounded shape correction. In particular,
this approximate soliton solution of Eq.~(\ref{pKdV}) reads (see, e.g., Ref.~\cite{mar3}):
\begin{eqnarray}
q(\xi,\tau ) = - \frac{\alpha }{{2{c^2}}}\left( {A\sech^2}\theta
+ \varepsilon {A^2}{\lambda_1}\sech^2\theta  + \varepsilon {A^2}{\lambda_2}\sech^4\theta \right) +O(\epsilon^2),
\label{hosol} \\
\theta  = \kappa \left(\xi + \frac{\alpha }{{8c}}V \tau - {\xi_0} \right),
\quad
V = 2A - \frac{{32c}}{\alpha }\varepsilon {c_4}{A^2},
\end{eqnarray}
where $A=2\kappa^2$, with $\kappa\in\mathbb{R}$ being the free $O(1)$ parameter of the
KdV soliton in Eq.~(\ref{kdvsoli}), while the constants $\lambda_1$ and $\lambda_2$ are given by:
\[
\hspace{-2cm}{\lambda_1} = \frac{{120{c^4}{c_4} + 2{c^2}\alpha {c_2} + 8{c^2}\alpha {c_3}
+ {\alpha ^2}}}{{3{c^3}\alpha }},\quad
{\lambda_2} = \frac{{ - 360{c^4}{c_4} - 6{c^2}\alpha {c_2} - 12{c^2}\alpha {c_3}
- {\alpha ^2}}}{{6{c^3}\alpha }}.
\]
Obviously, in the limit $\epsilon \rightarrow 0$ the soliton of Eq.~(\ref{hosol}) reduces to the KdV
soliton of Eq.~(\ref{kdvsoli}).

Furthermore, employing the perturbation theory for solitons \cite{km1,km2,yubo}, we may obtain the
following results. First, due to the presence of the perturbation $P(q)$ [see Eq.~(\ref{pofq})],
the parameter $\kappa$ becomes time-dependent, namely $\kappa \mapsto \kappa(t)$, featuring
an evolution determined by:
\begin{equation}
\frac{d\kappa}{dt} \propto -\frac{1}{4\kappa}
\int_{-\infty }^{+\infty }P(q_{s}){\rm sech}^{2}Z{\rm d}Z,
\label{koft}
\end{equation}
where $q_{s}$ is the soliton of the unperturbed KdV equation [see Eq.~(\ref{kdvsoli})].
Second, the amplitude of the radiation tails $q_{R}$ produced by the perturbation $P(q)$
is proportional to the factor $R$ given by:
\begin{equation}
q_R \propto R=\frac{1}{4\kappa^{5}}\int_{-\infty }^{+\infty }P(q_{s})\tanh^{2}Z{\rm d}Z.
\label{radi}
\end{equation}
Evidently, since $P(q_{s})$ is an odd function of $Z$ while ${\rm sech}^2Z$ and $\tanh^2Z$
are even functions of $Z$, both integrals in Eqs.~(\ref{koft}) and (\ref{radi}) vanish.
Thus, the parameter $\kappa$ which characterizes the soliton amplitude remains time-independent
and, at the same time, no radiation tails are produced in this higher-order KdV approximation. 

\section{Conclusions}

In conclusion, we have used multiscale expansion methods to study the dynamics of dark solitons
in weakly nonlocal media, governed by a nonlinear Schr{\"o}dinger model. In particular, we have
analyzed the hydrodynamic form of the model and considered at first the leading-order of 
approximation, where only the first moment of the medium's response function is present.
At an intermediate stage of the asymptotic analysis, we derived a Boussinesq/Benney-Luke 
(BBL) equation. Using a traveling wave ansatz, we derived exact solitary wave solutions of 
this equation, in the limiting case where the velocity of the solitary wave is sufficiently 
close to (and below) the speed of sound. Then, we considered the long-time behavior of the 
BBL equation and, upon introducing relevant scales and asymptotic expansions, we reduced the 
BBL model to a pair of Korteweg-de Vries (KdV) equations that govern right- and left-propagating 
waves. We have also shown that if the formal perturbation parameter becomes sufficiently small 
then the BBL solitary wave transforms into the KdV soliton.

We also used the reductive perturbation method to analyze higher-order
effects. We thus considered the model at the next order of approximation, where the second moment
of the response function comes into play. In this case, we found that dark solitons are governed
by a perturbed KdV (pKdV) equation which, as it has been shown in the past, can be approximated by a
higher-order integrable system. We have presented the exact soliton solution of the pKdV equation, 
and employed the perturbation theory for solitons to show that the soliton amplitude remains 
unchanged, while no radiation tails are produced during the evolution. Thus, it can be concluded 
that, in the presence of the higher-order effects, the dark soliton's shape and 
velocity is only insubstantially changed.

Our analysis and results suggest interesting directions for future investigations. First, it would
be relevant to study analytically the dynamics of the derived soliton solutions in a 
higher-dimensional setting, and investigate the role of weak nonlocality on the transverse 
modulational instability of dark solitons (see, e.g., Ref.~\cite{trillo} for a relevant study). 
It would also be relevant to study dispersive shock waves, and particularly the role of 
higher-order effects, in weakly nonlocal media. Naturally, exploring numerically the
quantitative aspects of the predictions herein both on the original
dynamical system of the weak nonlocality, as well as in the full
original setting of the model featuring the nonlocal kernel, would be of
particular relevance and interest. Various predictions including the existence of
the antidark, supersonic solitons or the range over which the gray
subsonic solitons may exist are of particular interest within such an investigation.
Finally, the derivation of higher-order nonlinear evolution equations that describe 
effectively soliton dynamics in higher-dimensional, fully nonlocal media is another 
quite interesting theme. Pertinent studies are in progress and relevant results 
will be reported elsewhere.

\end{document}